\documentclass[12pt]{article}
\usepackage{epsfig}
\voffset= - 0.5in
\hoffset= - 0.6in         
\def\lsim{\mathrel{\rlap{\lower3pt\hbox{\hskip0pt$\sim$}}
    \raise1pt\hbox{$<$}}}
\def\gsim{\mathrel{\rlap{\lower4pt\hbox{\hskip1pt$\sim$}}
    \raise1pt\hbox{$>$}}}

\catcode`\@=11
\def\marginnote#1{}

\newcount\hour
\newcount\minute
\newtoks\amorpm
\hour=\time\divide\hour by60
\minute=\time{\multiply\hour by60 \global\advance\minute by-\hour}
\edef\standardtime{{\ifnum\hour<12 \global\amorpm={am}%
        \else\global\amorpm={pm}\advance\hour by-12 \fi
        \ifnum\hour=0 \hour=12 \fi
        \number\hour:\ifnum\minute<10 0\fi\number\minute\the\amorpm}}
\edef\militarytime{\number\hour:\ifnum\minute<10 0\fi\number\minute}

\def\draftlabel#1{{\@bsphack\if@filesw {\let\thepage\relax
   \xdef\@gtempa{\write\@auxout{\string
      \newlabel{#1}{{\@currentlabel}{\thepage}}}}}\@gtempa
   \if@nobreak \ifvmode\nobreak\fi\fi\fi\@esphack}
        \gdef\@eqnlabel{#1}}
\def\@eqnlabel{}
\def\@vacuum{}
\def\draftmarginnote#1{\marginpar{\raggedright\scriptsize\tt#1}}

\def\draft{\oddsidemargin -.5truein
        \def\@oddfoot{\sl preliminary draft \hfil
        \rm\thepage\hfil\sl\today\quad\militarytime}
        \let\@evenfoot\@oddfoot \overfullrule 3pt
        \let\label=\draftlabel
        \let\marginnote=\draftmarginnote
   \def\@eqnnum{(\theequation)\rlap{\kern\marginparsep\tt\@eqnlabel}%
\global\let\@eqnlabel\@vacuum}  }

\def\d{\partial}

\def\bea{\begin{eqnarray}}
\def\eea{\end{eqnarray}}

\def\beq{\begin{equation}}
\def\eeq{\end{equation}}
\def\ba{\beq\new\begin{array}{c}}
\def\ea{\end{array}\eeq}
\def\be{\ba}
\def\ee{\ea}
\def\stackreb#1#2{\mathrel{\mathop{#2}\limits_{#1}}}
\def\Tr{{\rm Tr}}

\makeatletter
\newdimen\normalarrayskip              
\newdimen\minarrayskip                 
\normalarrayskip\baselineskip
\minarrayskip\jot
\newif\ifold             \oldtrue            \def\new{\oldfalse}
\def\arraymode{\ifold\relax\else\displaystyle\fi} 
\def\eqnumphantom{\phantom{(\theequation)}}     
\def\@arrayskip{\ifold\baselineskip\z@\lineskip\z@
     \else
     \baselineskip\minarrayskip\lineskip2\minarrayskip\fi}
\def\@arrayclassz{\ifcase \@lastchclass \@acolampacol \or
\@ampacol \or \or \or \@addamp \or
   \@acolampacol \or \@firstampfalse \@acol \fi
\edef\@preamble{\@preamble
  \ifcase \@chnum
     \hfil$\relax\arraymode\@sharp$\hfil
     \or $\relax\arraymode\@sharp$\hfil
     \or \hfil$\relax\arraymode\@sharp$\fi}}
\def\@array[#1]#2{\setbox\@arstrutbox=\hbox{\vrule
     height\arraystretch \ht\strutbox
     depth\arraystretch \dp\strutbox
     width\z@}\@mkpream{#2}\edef\@preamble{\halign
\noexpand\@halignto
\bgroup \tabskip\z@ \@arstrut \@preamble \tabskip\z@ \cr}%
\let\@startpbox\@@startpbox \let\@endpbox\@@endpbox
  \if #1t\vtop \else \if#1b\vbox \else \vcenter \fi\fi
  \bgroup \let\par\relax
  \let\@sharp##\let\protect\relax
  \@arrayskip\@preamble}
\def\eqnarray{\stepcounter{equation}%
              \let\@currentlabel=\theequation
              \global\@eqnswtrue
              \global\@eqcnt\z@
              \tabskip\@centering
              \let\\=\@eqncr
 \halign to \displaywidth\bgroup
    \eqnumphantom\@eqnsel\hskip\@centering
    $\displaystyle \tabskip\z@ {##}$%
    \global\@eqcnt\@ne \hskip 2\arraycolsep
         $\displaystyle\arraymode{##}$\hfil
    \global\@eqcnt\tw@ \hskip 2\arraycolsep
         $\displaystyle\tabskip\z@{##}$\hfil
         \tabskip\@centering
    &{##}\tabskip\z@\cr}
\begingroup\ifx\undefined\newsymbol \else\def\input#1 {\endgroup}\fi
\font\upright=cmu10 scaled\magstep1
\def\stroke{\vrule height8pt width0.4pt depth-0.1pt}
\def\topfleck{\vrule height8pt width0.5pt depth-5.9pt}
\def\botfleck{\vrule height2pt width0.5pt depth0.1pt}
\def\Zmath{\vcenter{\hbox{\numbers\rlap{\rlap{Z}\kern 0.8pt\topfleck}\kern
2.2pt
                   \rlap Z\kern 6pt\botfleck\kern 1pt}}}
\def\Qmath{\vcenter{\hbox{\upright\rlap{\rlap{Q}\kern
                   3.8pt\stroke}\phantom{Q}}}}
\def\Nmath{\vcenter{\hbox{\upright\rlap{I}\kern 1.7pt N}}}
\def\Cmath{\vcenter{\hbox{\upright\rlap{\rlap{C}\kern
                   3.8pt\stroke}\phantom{C}}}}
\def\Rmath{\vcenter{\hbox{\upright\rlap{I}\kern 1.7pt R}}}
\def\Z{\ifmmode\Zmath\else$\Zmath$\fi}
\def\Q{\ifmmode\Qmath\else$\Qmath$\fi}
\def\N{\ifmmode\Nmath\else$\Nmath$\fi}
\def\C{\ifmmode\Cmath\else$\Cmath$\fi}
\def\R{\ifmmode\Rmath\else$\Rmath$\fi}

\def\stackreb#1#2{\mathrel{\mathop{#2}\limits_{#1}}}
\def\Tr{{\rm Tr}}

\def\d{\partial}

\def\2{{1\over 2}}
\def\N2{${\cal N}=2$}
\def\4N{${\cal N}=4$}
\def\1N{${\cal N}=1$}

\def\ba{\beq\new\begin{array}{c}}
\def\ea{\end{array}\eeq}
\def\be{\ba}
\def\ee{\ea}
\def\stackreb#1#2{\mathrel{\mathop{#2}\limits_{#1}}}

\begin{document}
\def\half{{\textstyle{1\over2}}}
\newcommand{\qt}{\tilde q}
\newcommand{\E}{{\cal E}}
\newcommand{\qtu}{\tilde q_{1}}
\newcommand{\qtd}{\tilde q_{2}}
\newcommand{\ntwo}{${\cal N}=2\;$}
\newcommand{\none}{${\cal N}=1\;$}
\newcommand{\nfour}{${\cal N}=4\;$}
\newcommand{\vp}{\varphi}
\newcommand{\ptl}{\partial}
\renewcommand{\theequation}{\thesection.\arabic{equation}}
\newcommand{\rf}[1]{(\ref{#1})}

\begin{titlepage}
\renewcommand{\thefootnote}{\fnsymbol{footnote}}

\begin{flushright}
FIAN/TD-07/04\\
ITEP/TH-33/04
\end{flushright}

\vspace{1.3cm}

\begin{center}
\baselineskip20pt
{\bf \Large Non-Abelian Vortices in N=1* Gauge Theory}
\end{center}

\begin{center}

\vspace{0.5cm}

{\large
{ \bf   V.~Markov$^{a,d}$ },
{ \bf A.~Marshakov$^{b,c}$} and { \bf A.~Yung$^{a,c}$}}

\vspace{0.5cm}

$^a${\it Petersburg Nuclear Physics Institute, Gatchina, Russia}\\

$^b${\it Theory Department, Lebedev Physics Institute,
 Moscow, Russia}\\

$^c${\it Institute of Theoretical
and Experimental Physics, Moscow, Russia}\\

$^d${\it Helmholtz Institut f\"ur Strahen und Kernphysik, Bonn
University, Germany}  \\

\vfil

\end{center}

\begin{center}
{\large\bf Abstract} \vspace*{.2cm}
\end{center}

\begin{quotation}

\noindent
We consider the ${\cal N}=1^{*}$ supersymmetric $SU(2)$ gauge theory
and demonstrate that the $Z_2$ vortices in this theory acquire orientational
zero modes, associated with the rotation of magnetic flux inside $SU(2)$
group, and turn into the non-Abelian strings, when the
masses of all chiral fields become equal. These non-Abelian strings are
not BPS-saturated. We study the effective theory on the string
world sheet and show that it is given by two-dimensional
non-supersymmetric $O(3)$ sigma model. The confined 't Hooft-Polyakov monopole
is seen as a junction of the $Z_2$-string and anti-string, and as a kink
in the effective world sheet sigma model. We calculate its mass and
show that besides the four-dimensional confinement of monopoles,
they are also confined in the two-dimensional theory: the monopoles stick to
anti-monopoles to form the meson-like configurations on the strings
they are attached to.

\end{quotation}

\end{titlepage}


\setcounter{footnote}{0}
\setcounter{equation}{0}

\section{Introduction}

The idea of confinement as a dual
Meissner effect upon condensation   of    monopoles
was suggested many years ago by  't Hooft,  Mandelstam and
 Polyakov \cite{HMP}.
Understanding of the  electomagnetic
duality in \ntwo supersymmetric gauge theories  allowed Seiberg
and Witten to present the quantitative description of this
phenomenon \cite{SW1,SW2}.

Recall, first, the basic idea of the confinement mechanism \cite{HMP}.
Once magnetic (electric) charges condense, the electric
(magnetic) flux is confined in the Abrikosov-Nielsen-Olesen (ANO)
flux tube \cite{ANO} connecting heavy trial electric (magnetic)
charge and anti-charge.  The energy of the ANO string increases
lineary with its length. This ensures the linear increasing
confining potential between the heavy electric (magnetic) charges
and anti-charges.

Later studies of confinement in \ntwo QCD showed
\cite{DS,HSZ,VY,S,Yrev} that generally confinement in
this theory is essentially Abelian.
At the first stage, the gauge group is broken down to an
Abelian subgroup at larger scale by VEV's of the adjoint scalars.
Then the Abelian subgroup is broken down completely (or to
its discrete subgroup) at much smaller energy scale
by condensation of the quarks or monopoles. At
the second stage the Abelian ANO flux tubes are formed leading to
the confinement of magnetic or electric charges respectively.

However if one thinks of understanding the confinement mechanism in
real QCD or in \none gauge theories, a somewhat different pattern
of the gauge symmetry breaking without this "Abelization" is
probably needed \cite{MY}. In particular,
the flux tubes should have some non-Abelian features, which were
recently found in \cite{ABEKY} in the context of
\ntwo QCD with the gauge group $SU(2)\times U(1)$ with two
flavors of quarks, perturbed by the Fayet-Iliopoulos (FI)
term \cite{FI} of the $U(1)$ factor, (more
generally $SU(N)\times U(1)$ gauge group with $N$ flavors of
quarks, see also \cite{SYmon}; similar results  in three
dimensions were obtained in \cite{HT1}).
At large values of the quark masses the vacuum where both
flavors of squarks condense is in the weak coupling regime, and
quasiclassical analysis is applicable. Flux tubes in this vacuum
were found in \cite{MY}, and
it was demonstrated in \cite{ABEKY} that
in the limit of equal quark masses
the diagonal $SU(2)$ subgroup of global gauge and flavor groups
(called $SU(2)_{C+F}$) remains unbroken both by condensates of adjoint
and fundamental scalars. It turns out, that existence of this unbroken
subgroup ensures, that elementary BPS flux tubes acquire the
orientational
modes, associated with the rotation of the color
magnetic flux, and these modes make the strings to be
non-Abelian \footnote{Recently similar model for non-Abelian
strings in six dimensions was presented in \cite{ENS}.}

Let us note that the flux tubes in non-Abelian theories at weak
coupling  were studied in numerous papers in recent years
\cite{VS,HV,Su,SS,KB,KoS}. In particular, the $Z_N$ strings
associated with the center of gauge group were constructed.
However, in all these constructions the flux was
always directed along a fixed vector (in the
Cartan subalgebra), and no moduli which could govern its
orientation in the group space were ever found.

In this paper we suggest the simpler model for non-Abelian
strings (compare to \cite{ABEKY,HT1}).
We consider the \1N$^*$ supersymmetric $SU(2)$ gauge theory, or the
\nfour theory with the
mass terms for three \1N chiral superfields, and we take two
equal masses, say $m_1=m_2= m$, while the third mass $m_3$
is generically distinct. The \nfour
supersymmetry is the broken down to \1N, unless $m_3=0$ when
the theory has \ntwo supersymmetry and becomes \ntwo gauge theory with
adjoint matter (more strictly two
flavors of adjoint matter with equal masses).

Classically the vacuum structure of this theory was studied in \cite{VW},
while the quantum effects were taken into account in \cite{DW},
using the parallels between the Seiberg-Witten theories and integrable
systems \cite{GKMMM}, of which the deformed \nfour theory together with
the integrable Calogero-Moser family is the most elegant example (see also
\cite{SWCa}).
The \nfour theory with $SU(2)$ gauge
group has three vacua, for small coupling\footnote{Note that the coupling of
unbroken \nfour theory $g$ does not run since the theory
is conformal.} $g\ll 1$ one of these vacua is in the weak coupling.
All three adjoint scalars condense in this vacuum, therefore it
is called Higgs vacuum  \cite{VW,DW}. Other two vacua of the
theory are always in the strong coupling, for small
$m_3$ these are the monopole and dyon vacua of the
perturbed \ntwo theory \cite{SW1}.

In this paper we concentrate
on the Higgs vacuum in the weak coupling regime.
In this vacuum the gauge group $SU(2)$ is broken down
to $Z_2$ by the adjoint scalar VEV's, therefore there are stable $Z_2$
non-BPS flux tubes associated with the $\pi_1(SU(2)/Z_2)=Z_2$.
Note, that as soon as electric (color) charges develop
VEV's, the $Z_2$ strings carry magnetic fluxes and give rise to the
confinement of monopoles. In the $SU(2)$ theory there is a 't Hooft-Polyakov
monopole \cite{tHP} with a unit magnetic charge.
Since the $Z_2$-string's charge is $1/2$, it cannot end on a
monopole (and this is the reason for the
stability of $Z_2$ strings). Instead, the confined monopole appears to be a
{\em junction} of the $Z_2$-string and anti-string (monopoles as string
junctions were considered earlier in \cite{HiKi,EA,T} and
recently as junctions of the non-Abelian BPS strings in
\cite{SYmon,HT2}).  Note that it was shown in
\cite{Bais,PrVi,MY,Kn,ABEK} that in different models the monopole
fluxes match those of the flux tubes, hence the monopoles can be confined by
one or several strings.

We show that at the special value $m_3=m$ there is a
diagonal $O(3)_{C+F}$ subgroup of the global gauge group $SU(2)$
and flavor $O(3)$ group, unbroken by vacuum condensates.
Like in \cite{ABEKY,HT1}, the presence
of this group leads to emergence of orientational zero modes
of the $Z_2$-strings associated with rotation of the color magnetic
flux of a string inside the $SU(2)$ gauge group, which makes a $Z_2$
string genuinely non-Abelian\footnote{Note that
non-translational zero modes of string were considered
earlier in \cite{Wss,Hi,ABCMW}. In particular, the modes
considered in \cite{ABCMW} are somewhat similar to our
orientational zero modes. However, the difference is that
unbroken symmetry in \cite{ABCMW} is gauged. Thus, in contrast to our case,
there are massless gauge fields
in the bulk, which creates certain problems with normalizability
of the zero modes.}.

Next, we derive the two-dimensional effective theory for the orientational
zero modes on the string world sheet. It turn out to be
(a non-supersymmetric!) two-dimensional $O(3)$ sigma model~\footnote{A 
three-dimensional $O(3)$ sigma model in the context of
"superconductivity" for the Yang-Mills theory with two flavors was
studied recently in \cite{NiHo}.}.
Note, that effective world
sheet theory for the BPS non-Abelian strings, considered in
\cite{HT1,ABEKY,T,SYmon,HT2}, is the \ntwo SUSY $O(3)$ sigma model
(or the \ntwo SUSY $CP^{N-1}$ model for the gauge group
$SU(N)\times U(1)$). Here we
deal with the {\em non-BPS} strings therefore the effective world sheet
theory is not supersymmetric (to be called just $O(3)$ sigma model
in what follows). We also consider the case when
$m_3$ is not exactly equal to $m$, with
$\Delta m \equiv |m_3-m|\ll m$. In this case a shallow potential
in the sigma model
is generated, which makes it classically massive.

Then we discuss the confined monopole or a junction of
non-Abelian $Z_2$-strings. We identify this monopole as an $O(3)$
sigma model kink \cite{T,SYmon,HT2} and calculate its mass.
Classically the mass of this monopole vanishes in the limit
$\Delta m\to 0$, while its size become infinite
(cf. with \cite{We}). We show, however, that this
does not happen in quantum theory. When the non-perturbative effects
in the $O(3)$ sigma model are taken into account,
the monopole mass  is  determined by the dynamical scale of the
$O(3)$ sigma model $\Lambda_{\sigma}$, and its size remains finite
(of order of $\Lambda_{\sigma}^{-1}$).

Finally, we demonstrate that besides the four-dimensional
confinement, which ensures
that the monopoles are attached to the strings, they are also
confined in the two-dimensional sense. Namely, the monopoles stick to
the anti-monopoles on the string
they are attached to, and form a meson-like configuration.

The paper is organized as follows. In Sect.~\ref{ss:model} we review the
\1N$^{*}$ gauge theory near the Higgs vacuum. In
Sect.~\ref{ss:astrings} we construct solutions for the Abelian
$Z_2$-strings. First, we consider the Abelian limit of small
$m_3$, when the theory has \ntwo supersymmetry and possesses the
whole tower of ANO Abelian BPS strings. When we increase $m_3$,
all strings with multiple winding numbers become unstable, and we
are left with lightest stable $Z_2$-strings. In
Sect.~\ref{ss:nastrings} we take the limit $m_3=m$ and construct
the orientational zero modes of the non-Abelian $Z_2$-strings.
In Sect.~\ref{ss:ws}  we derive the effective world sheet theory
for these strings and
in Sect.~\ref{ss:dynamics} consider its dynamics. Sect.~\ref{ss:conclusion}
contains our conclusions.

\section{The model
\label{ss:model}}

In terms of \none supermultiplets, the \nfour supersymmetric
gauge theory with the $SU(2)$ gauge group contains a vector
multiplet, consisting of the gauge field $A_{\mu}^a$ and gaugino
$\lambda^{\alpha a}$, and three chiral multiplets $\Phi^a_A$, $A=1,2,3$,
all in the adjoint representation of the gauge group, with $a=1,2,3$ being
the $SU(2)$ color index. The superpotential of the \nfour gauge theory
reads
\be
W_{{\cal N}=4}= -\frac{\sqrt{2}}{g^2}
\varepsilon_{abc}\Phi_1^a\Phi_2^b\Phi_3^c.
\label{n4sup}
\ee
One can deform this theory, breaking \nfour supersymmetry down to \N2,
by adding the mass terms
with equal masses $m$, say for the first two flavors of the adjoint matter
\be
\label{N2mass}
W_{{\cal N}=2} = {m\over 2g^2}\sum_{A=1,2} \left(\Phi_A^a\right)^2
\ee
Then the third flavor combines with the vector
multiplet to form a \ntwo vector supermultiplet, while the first two
flavors \rf{N2mass} can be treated as \N2 massive adjoint matter.
One can further break supersymmetry down to \1N, adding a mass
term  to the $\Phi_3$ multiplet
\be
\label{N1mass}
W_{{\cal N}=1^*} = {m_3\over 2g^2} \left(\Phi_3^a\right)^2
\ee
Then bosonic part of the action
reads
\be
S_{{\cal N}=1^*} = {1\over g^2}\int d^4 x\left(
\frac14\left(F_{\mu\nu}^a\right)^2 +
\sum_A \left| D^{\rm
adj}_\mu\ \Phi_A^a\right|^2
+\frac12\sum_{A,B}\left[(\bar{\Phi}_A\bar{\Phi}_B)(\Phi_A\Phi_B)-
(\bar{\Phi}_A\Phi_B)(\bar{\Phi}_B\Phi_A)\right]+\right.
\\
\left.
+\sum_{A}\left|\frac1{\sqrt{2}}
\varepsilon_{abc}\varepsilon^{ABC}\Phi^b_B\Phi^c_C
-m_A\Phi^a_A\right|^2\right) +\  {\rm fermionic\ \ terms},
\label{su2}
\ee
where $F_{\mu\nu}^a = \d_\mu A_\nu^a-\d_\nu A_\mu^a+
\varepsilon^{abc}A_\mu^bA_\nu^c$, $D^{\rm
adj}_\mu\ \Phi_A^a = \d_\mu \Phi^a_A + \varepsilon^{abc}A_\mu^b\Phi^c_A$,
and we use the same notations $\Phi^a_A$ for the scalar components
of the corresponding chiral superfields.

In this paper we are going to study the so called Higgs vacuum of the
theory \rf{su2}, when all three adjoint scalars develop the
VEV's of the order of $m, \sqrt{mm_3}$.
 The scalar condensates $\Phi^a_A$ can be
written in the form of the following 3$\times$3
color-flavor matrix (convenient
for the $SU(2)$ gauge group and three \nfour
"flavors")
\be
\label{gvac}
\langle\Phi_A^a\rangle =\frac{1}{\sqrt{2}}\left(
\begin{array}{ccc}
  \sqrt{mm_3} &
 0& 0 \\
 0 &
\sqrt{mm_3} & 0 \\
  0 & 0 & m
\end{array}
\right),
\ee
and these VEV's break completely the $SU(2)$ gauge group. The masses
of the W-bosons $A_{\mu}^{1,2}$ are
\be
m^2_{1,2}=m^2 +mm_3,
\label{Wmass}
\ee
while the mass of the "photon" $A_{\mu}^3$ is
\be
m^2_{\gamma}=2mm_3.
\label{phmass}
\ee
In what follows,
we will be especially interested in a particular point
of the parameter space of the theory where $m_3=m$.
For this value of $m_3$, (\ref{gvac}) turns to be a
symmetric color-flavor locked \cite{BarH} vacuum
\be
\label{svac}
\langle\Phi_A^a\rangle =\frac{m}{\sqrt{2}}\left(
\begin{array}{ccc}
  1 &
 0& 0 \\
 0 &
1 & 0 \\
  0 & 0 & 1
\end{array}
\right),
\ee
This symmetric vacuum respect global $O(3)_{C+F}$ symmetry
\be
\Phi\rightarrow O\Phi O^{-1},\;\;\; A_{\mu}^a\rightarrow
O^{ab}A_{\mu}^b,
 \label{c+f}
\ee
which combine transformations from the
global color and flavor groups. As we see later, it is this symmetry
that is responsible for the presence of non-Abelian strings in
the vacuum (\ref{svac}).

Now let us study the mass spectrum of the theory around the vacuum
(\ref{svac}). From (\ref{Wmass}) and (\ref{phmass}) we see that
masses of all gauge bosons are equal and given by
\be
m^2_{g}=2m^2.
\label{gmass}
\ee
This means, in particular, that at the point $m_3=m$ we loose all  traces of
the "Abelization" in our theory present at other values of $m_3$, see
more details below.

To find masses of adjoint scalars consider the mass matrix coming
from (\ref{su2}). For $m_3=m$ its 18 eigenvectors
(if count the real degrees of freedom) are combined as follows: 3
states have zero mass,
these states are "eaten" by the Higgs mechanism; other 3 states have
the same mass (\ref{gmass}) as the gauge bosons. They become
superpartners of the gauge bosons in 3 massive vector \none
supermultiplets. Other 6 (=3$\times$2) states acquire the masses
\be
m^2_{1}=m^2,
\label{1mass}
\ee
while the remaining 6 states acquire the mass
\be
m^2_{2}=4m^2.
\label{2mass}
\ee
These states are the scalar components of the 3+3 chiral \none
supermultiplets with the masses (\ref{1mass}) and (\ref{2mass}),
the factor 3 everywhere stands for the color multiplicity
(since the states come in triplets of the unbroken at $m_3=m$ color-flavor
symmetry (\ref{c+f})).

To conclude this section let us note that the coupling
$g$ in (\ref{su2}) is \nfour coupling  constant.
It does not run in \nfour theory at scales above $m$
and we take it to be small
\be
g\ll 1.
\label{n4coupling}
\ee
At the scale $m$ the gauge group $SU(2)$ is broken
in the vacuum (\ref{svac}) by the scalar VEV's. As we see
later, the running of the coupling constant below scale $m$
is determined by the beta-function of the effective
two-dimensional sigma model.

\section{Abelian  strings
\label{ss:astrings}}
\setcounter{equation}{0}

In this section we consider strings or magnetic flux tubes
in the model (\ref{su2}). We start with the limit of small
$m_3$ in which the low energy effective theory of (\ref{su2})
is \ntwo QED and consider Abelian ANO strings \cite{ANO} in this theory. Then
we increase $m_3$ and embed the ANO strings into the full
non-Abelian theory (\ref{su2}). We will find that only
the Abelian strings with minimal
winding number (the $Z_2$ strings) remain stable
in the full theory (\ref{su2}).

\subsection{U(1) truncation and \ntwo limit}

In fact in the limit of small $m_3$ the mass term  (\ref{N1mass})
does not break \ntwo
supersymmetry \cite{HSZ,VY}, the model reduces to \ntwo QED with
a FI term. At $m_3=0$ the theory has a Coulomb branch parameterized
by arbitrary VEV of $\Phi^a_3$ which by gauge rotation can
be directed, say, along the third axis in color space,
$\langle \Phi^a_3\rangle =\delta^{a3}\langle a\rangle$.
Thus the $SU(2)$ group is broken down to
$U(1)$ and the theory becomes essentially Abelian.

The Coulomb branch has singular points
where some   matter adjoint fields or monopoles or dyons become massless
\cite{SW1,SW2}.  These singular points become isolated vacua
once small parameter $m_3$ is introduced.

In this paper we will concentrate at the singular point
in
which adjoint matter becomes massless.
At non-zero $m_3$ this point corresponds to \none Higgs vacuum
(\ref{gvac}).  At small \nfour coupling $g^2$
 this vacuum is in the weak coupling
regime. Let us work out the effective low energy theory in this
vacuum.  As soon as $SU(2)$ group is broken down to $U(1)$, the
W-boson supermultiplets become heavy. Thus our low energy theory
is \ntwo QED. It includes the third color component of the
gauge supermultiplet, neutral scalar $a$ as well as  its fermionic
superpartner.

Now let us see which matter fields become massless in
the Higgs  vacuum at $m_3=0$. To do this we analyzes the mass
matrix of the matter fields given by superpotentials
(\ref{n4sup}), (\ref{N2mass}). It is easy to check that
at the point $a=m/\sqrt{2}$ we have two eigenvectors with zero
mass, which can be parameterized as
\be
\label{light}
\left(
\begin{array}{ccc}
  1 &
0 & 0 \\
0 &
1 & 0 \\
  0 & 0 & 0
\end{array}\right) \;\;\;\; {\rm and} \;\;\;\;
\left(
\begin{array}{ccc}
  0 &
-1 & 0 \\
1 &
0 & 0 \\
  0 & 0 & 0
\end{array}\right),
\ee
in the $3\times3$ matrix notation \rf{gvac} for $\Phi^a_A$.

Thus our low energy effective QED besides $U(1)$ gauge multiplet
includes the following bosonic mater fields
\be
\label{phi}
\Phi_A^a =\left(
\begin{array}{ccc}
  {g\over 2}\left(\chi+\tilde\chi\right) &
-{g\over 2i}\left(\chi-\tilde\chi\right) & 0 \\
 {g\over 2i}\left(\chi-\tilde\chi\right) &
{g\over 2}\left(\chi+\tilde\chi\right) & 0 \\
  0 & 0 & a
\end{array}\right)
\ee
The normalization here is chosen to ensure standard kinetic terms
for the charged chiral fields   $\chi$ and $\tilde{\chi}$ which belong to
a matter hypermultiplet of \ntwo SUSY QED. The superpotential
(\ref{n4sup}), (\ref{N2mass}) now turns to be
\be
W_{QED}=-\sqrt{2}{\chi}a\tilde\chi +m{\chi}\tilde\chi.
\label{qedsup}
\ee
and we see that two flavors of adjoint matter
of \nfour theory form a single flavor of standard
charged hypermultiplet (with unit charge) in the low energy
effective \ntwo SQED.

Now let us restore a small mass term \rf{N1mass} for the $a$-field.
The bosonic action of the model acquires the form
\be
\label{qed}
S_{QED}=\int d^4 x\left({1\over 4g^2}(F_{\mu\nu}^3)^2 +
|D_\mu\chi|^2 + | D_\mu\bar{\tilde{\chi}}|^2 +
{1\over g^2}|\ptl_\mu a|^2 +
V(\chi,\tilde\chi,a)\right)
\ee
with the long derivative $D_\mu = \ptl_\mu - iA^3_\mu$ and potential
\be
\label{pot}
V(\chi,\tilde\chi,a) =
2g^2\left|\chi\tilde\chi -
\frac{m_3}{\sqrt{2}g^2} a\right|^2 + 2\left|a-\frac{m}{\sqrt{2}}
\right|^2\left(
|\chi|^2+|\tilde\chi|^2\right)
 +\frac{g^2}{2} \left(|\chi|^2-|\tilde{\chi}|^2\right)^2,
\ee
where the last contribution to the r.h.s. comes from the D-terms.

The vacuum of this theory is given by
\be
\langle a\rangle =\frac{m}{\sqrt{2}},
\\
\langle \chi\rangle =\langle \bar{\tilde{\chi}}\rangle
=\frac1{g}\sqrt{\frac{mm_3}{2}}.
\label{vac}
\ee
This model possesses the standard ANO $U(1)$ strings \cite{ANO}.

Consider, first, the "BPS case" when
$m_3\ll m$, then the mass term for the field
$a=\Phi_3$ reduces to the FI term \cite{VY} with the FI parameter
\be
\xi=\frac{mm_3}{g^2}.
\label{xi}
\ee
This limit corresponds to keeping only the constant term in the
expansion of $a$ near its VEV in the first term of the
potential (\ref{pot}), i.e.
\be
\left.\frac{m_3 a}{\sqrt{2}g^2}
\right|_{a =\frac{m}{\sqrt{2}}}\
\longrightarrow\ \frac{mm_3}{2g^2}=\frac{\xi}{2}.
\label{atoxi}
\ee
With this truncation the theory (\ref{qed}) is just (a bosonic part of)
\ntwo QED, and it has a BPS Abelian ANO string solutions. To find the string
with winding number $n=1$ (which we are interested in also in the non BPS
case below) one takes the standard ansatz
\be
\label{bpsstr}
\chi = \frac1{\sqrt{2}}\phi(r)e^{i\alpha},
\\
\tilde\chi = \frac1{\sqrt{2}}\phi(r)e^{-i\alpha},
\\
a=\frac{m}{\sqrt{2}},
\\
A^3_i = {\varepsilon_{ij}x_j\over r^2}\left(f(r)-1\right),\ \ \ \ i,j=1,2
\ee
where $(x_1,x_2)$ are co-ordinates in the plane orthogonal
to the string axis,
while $r=\sqrt{x_1^2+x_2^2}$ and $\alpha=\arctan (x_2/x_1)$ are
polar coordinates in this plane.
The profile functions in (\ref{bpsstr}) satisfy
the first order equations \cite{B}
\be
r\phi' - f\phi=0,
\\
\frac{1}{r}f' - g^2(\phi^2-\xi)=0,
\label{foe}
\ee
where primes denote derivatives with respect to $r$. These equations
 should be supplemented by the boundary conditions
\be
\phi(0) = 0, \ \ \ \ \phi(\infty) = \sqrt{\xi}
\\
f(0) = 1, \ \ \ \ f(\infty) = 0.
\label{bpsbc}
\ee
The tension of the BPS string with winding number $n=1$
is given by
\be
T_{BPS}= 2\pi\xi = \frac{2\pi mm_3}{g^2}.
\label{bpsten}
\ee
Strings in the limit of small $m_3$ of \1N$^{*}$ theory were
also studied in the last paper of ref.~\cite{Kn}.

\subsection{$Z_2$ strings
\label{ss:astringsz2}}

Now let us increase $m_3$, breaking \ntwo supersymmetry down to
\none. Still it is clear that the deformed theory has ANO strings,
however, they  are {\em no} longer BPS saturated; what happens is
that the short BPS string multiplet of \ntwo SUSY becomes a long
non-BPS string multiplet of \none theory \cite{VY}. The number of
states in the string multiplet remains the same
(2~bosonic + 2~fermionic).

As $m_3$ increases and reaches values of the order of $m$, the
fields from the QED Lagrangian (\ref{qed}) cannot be considered
as "light", eventually their masses ($\sim \sqrt{mm_3}$) become
of the same order as the masses of "heavy" non-Abelian fields.
Thus the QED description is no longer valid and we should study
the full non-Abelian theory (\ref{su2}). Nevertheless, one can
use the QED truncation (\ref{qed}) at the classical level to look
for the Abelian strings embedded into non-Abelian
theory\footnote{This is because the ansatz (\ref{phi}) is still
consistent with equations of motion.}.

The necessary modification now is to introduce a new profile
function for the field $a$, since we cannot use
(\ref{atoxi}) any longer, and field $a$ cannot remain
constant on the string
solution. The modified ansatz for the Abelian string
embedded in the non-Abelian theory is
\be
\label{str}
\chi = \frac1{\sqrt{2}}\phi(r)e^{i\alpha},
\\
\tilde\chi = \frac1{\sqrt{2}}\phi(r)e^{-i\alpha},
\\
a=a_0(r),
\\
A^3_i = {\varepsilon_{ij}x_j\over r^2}\left(f(r)-1\right),\ \ \ \ i,j=1,2
\ee
and the profile functions here satisfy now the non-BPS {\em second} order
differential equations
\be
\label{streq}
\phi''+{1\over r}\phi'-{1\over r^2}f^2\phi =
\phi\left(g^2\phi^2-\sqrt{2}m_3a_0\right)+2\phi\left(a_0-
\frac{m}{\sqrt{2}}\right)^2 ,
\\
a_0''+ {1\over r}a_0' = -\frac{m_3}{\sqrt{2}}\left(
g^2\phi^2-\sqrt{2}m_3 a_0\right)
+2g^2\phi^2\left(a_0-\frac{m}{\sqrt{2}}\right),
\\
f''- {1\over r}f' = 2g^2f\phi^2 .
\ee
with the boundary conditions
\be
\phi(0) = 0, \ \ \ \ \phi(\infty) = \sqrt{\xi}
\\
a_0'(0)=0, \ \ \ \ a_0(\infty) = \frac{m}{\sqrt{2}},
\\
f(0) = 1, \ \ \ \ f(\infty) = 0.
\label{bc}
\ee
The boundary conditions for the profile functions $\phi(r)$ and $f(r)$ are
standard for the Higgs and magnetic field on the vortex solution. The extra
profile function $a_0(r)$ saturates its VEV at $r=\infty$, while
the boundary condition $ a_0'(0)=0$ deserves a special comment.
Expanding equations (\ref{streq}) near the origin, one finds
two solutions for $a_0$ with the behavior $a_0\sim
{\rm const}$ and $a_0\sim \log{r}$ at $r\to 0$. The second solution
gives rise to an infinite tension and the condition $a_0'(0)=0$
selects the first one.

To get the string tension we substitute
the solution (\ref{str}) into the action (\ref{su2}), the result is
\be
T=2\pi\int_0^{\infty} rdr\left[
\frac{f'^2}{2g^2r^2} +\phi'^2+\frac{a_0'^2}{g^2}
+\frac{f^2\phi^2}{r^2}
+\frac{g^2}{2}\left(\phi^2-\frac{\sqrt{2}m_3a_0}{g^2}\right)^2
+2\phi^2\left(a_0-\frac{m}{\sqrt{2}}\right)^2\right]
\label{ten}
\ee
In the BPS limit $m_3 \ll m$ or $m_3\to 0$ with $\frac{mm_3}{g^2} = \xi ={\rm
fixed}$, eqs.~\rf{streq} possess a solution $a_0(r) = \frac{m}{\sqrt{2}}=
{\rm const}$, together with $\phi$ and $f$ satisfying \rf{foe}; the integral
\rf{ten} then can be easily calculated and gives $\left.T\right|_{m_3\to 0}
= 2\pi\xi f(0)$ which coincides with \rf{bpsten} due to \rf{bc}. We will
also see in Sect.~\ref{ss:numeric} that for small $m_3 \ll m$ the numeric
solutions to \rf{streq} do not differ too much from the solutions to the
 first order equations \rf{foe}.

It is worth noting that only the elementary strings with winding numbers
$n=\pm 1$ are stable when embedded in the non-Abelian theory. These are the
so called $Z_2$ strings (and anti-strings) with the {\em half}-charges in
the monopole units, associated with the center of the $SU(2)$ gauge group
 \be \pi_1(SO(3))=Z_2.
\label{top}
\ee
These strings correspond to winding around a semicircle
on the "equator" of the $SU(2)$ group space $S_3$, clearly this trajectory
with "fixed ends" cannot be shrunk to a point.

Instead all strings with multiple winding numbers
become unstable
at $m_3\sim m$, see, for example
 \cite{PrVi,SYmeta}.  Say, the string with winding number
$n=2$ correspond to winding along the whole equator on the $SU(2)$ sphere. This
trajectory can be shrunk to zero by contracting the loop
towards either north or south poles.

Finally, let us rewrite the string solution in the
singular gauge. In this gauge scalar fields do not wind at infinity
and the string flux come from the small circle around the string axis.
One has
\be
\label{sstr}
\chi =\bar{\tilde{\chi}}=\frac1{\sqrt{2}}\phi(r),
\ \ \ \ \
a=a_0(r),
\\
A^3_i = \frac{\varepsilon_{ij}x_j}{r^2}\, f(r),\ \ \ \ i,j=1,2.
\ee
This form of the string solution will be used in the next section.

\subsection{Large $m_3$ limit
\label{ss:lm3}}

Although we are mostly interested in this paper in the strings at
particular value $m_3=m$, it is rather instructive to consider also
the limit of large $m_3$, $m_3\gg m$. The Abelian string in \ntwo
QED (\ref{qed}) in the limit of $m_3\gg m$ were studied in detail in
\cite{VY}.

In the limit of large $m_3$ one can integrate out the field $a$ in
 (\ref{qed}), this results in the following scalar potential for
fields $\chi$ and $\tilde{\chi}$ (cf. with \cite{VY})
\be
\left. V(q,\tilde q)\right|_{m_3\to\infty}\  =\
\frac{g^2}2\left(|\chi|^2-|\tilde{\chi}|^2\right)^2 
+
{4g^4\over m_3^2}\left(|\chi|^2+|\tilde{\chi}|^2\right)\left|
\tilde{\chi}\chi-\frac{\xi}{2}\right|^2
\label{lm3pot}
\ee
This potential has a minimum at
$\langle\chi\rangle=\langle\tilde{\chi}\rangle=0$
with unbroken $U(1)$ gauge group as well as
that one from (\ref{vac}) with broken gauge group by $\langle\chi\rangle=
\langle\tilde\chi\rangle=\sqrt{\xi\over 2}$, and
below we concentrate only on the later one.

Calculating the $4\times4$ mass matrix, following from \rf{lm3pot}
near this vacuum, we get one zero eigenvalue (corresponding to the state
"eaten" by
the Higgs mechanism). Another one equals to $m_{\gamma}^2=2mm_3$,
(corresponding to the scalar superpartner of the photon (\ref{phmass})), while
two other eigenvalues are
\be
m_{H}^2=4m^2 .
\label{hmass}
\ee
and correspond to the mass of a single
\none chiral multiplet, containing two real scalars.

Consider now the ANO vortex in the QED with potential
(\ref{lm3pot}). From symmetry between $\chi$ and $\tilde{\chi}$ it
is clear that the classical solution can be found, using the ansatz
(\ref{sstr}) with a single complex scalar $\phi$.
The mass of the Higgs scalar $\phi$
is given by (\ref{hmass}).
At $m_3\gg m$ we have
\be
m_H \ll m_{\gamma} .
\ee
This condition means that we got the case of the extreme
type I superconductor with the Higgs mass much less then the mass
of the photon. Strings in this limit are studied in
\cite{Y99,VY}. It turns out that the main contribution to the
string tension comes from the logarithmically wide region
of intermediate $r$, $m_{\gamma}^{-1}\ll r \ll m_H^{-1}$. In this
region the Higgs field is essentially free and
has typical two-dimensional solution with logarithmic behavior.

The result for the string tension
with minimal winding  is \cite{Y99}
\be
T_I
= \frac{2\pi\xi}{\log{m_{\gamma}\over m_H}} =
\frac{T_{BPS}}{\log{m_{\gamma}\over m_H}},
\label{typeIten}
\ee
which comes from the kinetic
energy of the scalar field ("surface" energy).  The details of
the scalar potential become essential only in the region
$r\sim {1\over m_H}$, but the "volume"
energy coming from this region is suppressed by extra powers of
$\log {m_{\gamma}\over m_H}$ as compared with the one in
(\ref{typeIten}), see \cite{Y99,VY}.

Hence, we conclude that at large $m_3\gg m$ our effective QED behaves
as a type I superconductor and the string tension of the ANO vortex
is given by
\be
T_I\ \stackreb{{m_3\to\infty}}{\simeq}\
\frac{4\pi}{g^2}\ \frac{ mm_3}{\log{m_3\over 2m}},
\label{lm3ten}
\ee
where we have expressed the FI parameter $\xi$ and particle masses in
terms of $m$ and $m_3$
using (\ref{xi}), (\ref{phmass}) and (\ref{hmass}).  In the next section
we compare the numeric result for the
string tension (\ref{ten}) with (\ref{lm3ten}) at large $m_3$.

\subsection{Numeric solutions
\label{ss:numeric}}

In this section we discuss the numeric solutions to
the equations for the string profile
functions (\ref{streq}).


\begin{figure}[tp]
\epsfysize=9cm
\centerline{\epsfig{file=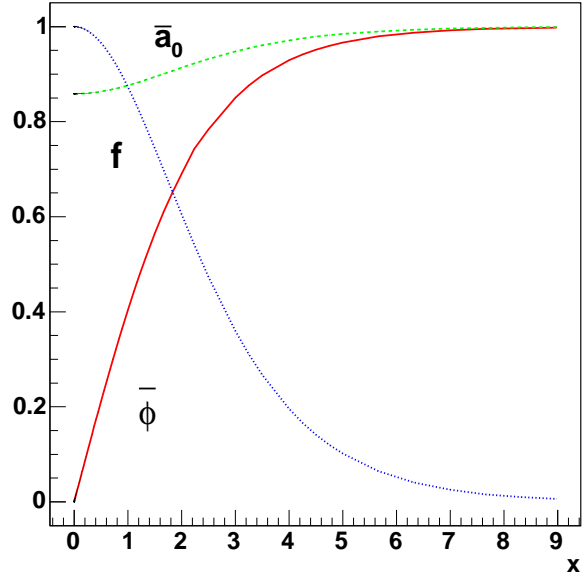,width=90mm,angle=0}}
\centerline{\epsfig{file=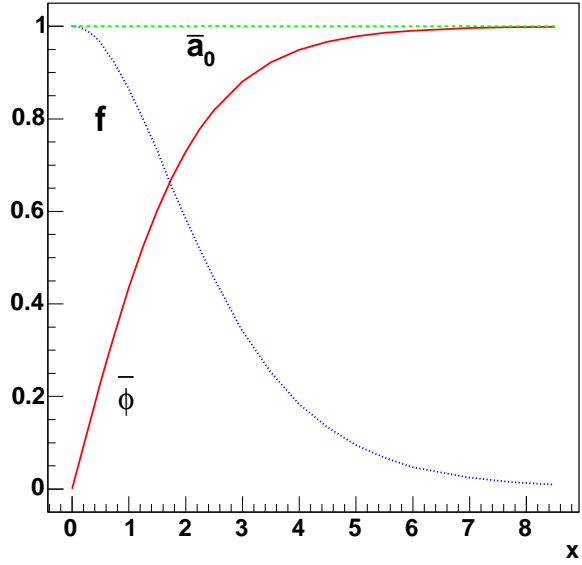,width=90mm,angle=0}}
\caption{Numeric solutions at small $\mu=m_3/m$. At the top picture we have
three non-BPS (normalized) profile functions
$\bar\phi (x)$, $\bar a_0(x)$ and $f(x)$ calculated numerically
for $\mu=0.3$. For comparison we present (at the bottom picture)
the BPS profile functions
(at $\mu=0$) for $\bar\phi (x)$ and $f(x)$, together with $\bar a_0(x)
\equiv 1$.}
\label{fi:m302}
\end{figure}
\begin{figure}[tb]
\epsfysize=9cm
\centerline{\epsfbox{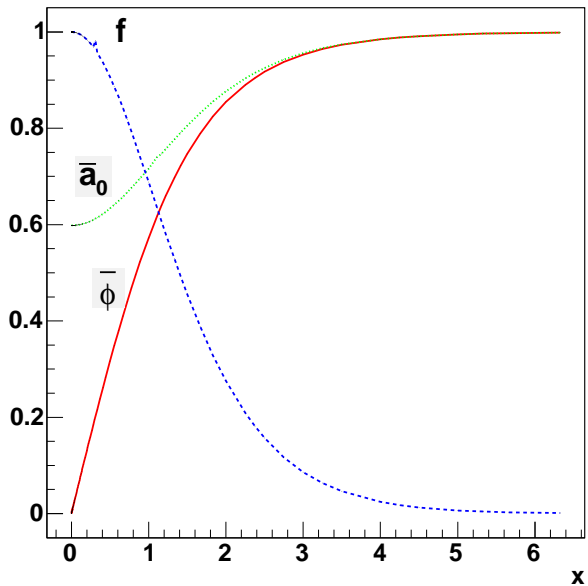}}
\caption{Numeric solutions for the profile functions at the non-Abelian
point $m_3=m$ ($\mu=1$)
for the functions $\bar\phi (x)$, $\bar a_0(x)$, and $f(x)$.}
\label{fi:m31}
\end{figure}

To solve the system of differential equations (\ref{streq}) with the
boundary conditions imposed at two points (\ref{bc}), we have used
a variable order and a variable step size within the
finite difference method with deferred
corrections \cite{Ascher}. Global error estimates were
produced to control the computation.

To simplify numeric calculations we have also introduced the
normalized (to their VEV's, i.e. $\bar{\phi},\bar{a}_0\to 1$
at $r\to \infty$) profile functions
\be
\bar{\phi} = \frac{g\phi}{\sqrt{mm_3}}, \;\;\;
\bar{a}_0 = \frac{\sqrt{2}a_0}{m}.
\label{npf}
\ee
of the dimensionless variable
\be
x=mr
\label{x}.
\ee
In this variables the string equations (\ref{streq}) have the only
remaining free dimensionless parameter $\mu = {m_3\over m}$; and
we study the
profile functions and string tension at different values of this
parameter.

To check our methods we started with the limit of small $\mu = {m_3\over m}$.
For  $\mu=0.3$  we plot solutions of  non-BPS  equations
(\ref{streq})
together with solutions of  BPS equations (\ref{foe}) at
fig.~\ref{fi:m302}
as functions of the $x$-variable  (\ref{x}).
One can see that the profile functions for the non-BPS $Z_2$-string
are indeed very close to those for the BPS string. In particular,
$\bar{a}_0$ does not go much away from unity
(note that for the BPS string $\bar{a}_0\equiv 1$).
However, we are mostly interested to study the strings at
$m_3=m$ or $\mu=1$.
Solutions for profile functions for this case are shown
at fig.~\ref{fi:m31}. We see that they are already essentially
different from the BPS case solutions.


\begin{figure}[tb]
\centerline{\epsfig{file=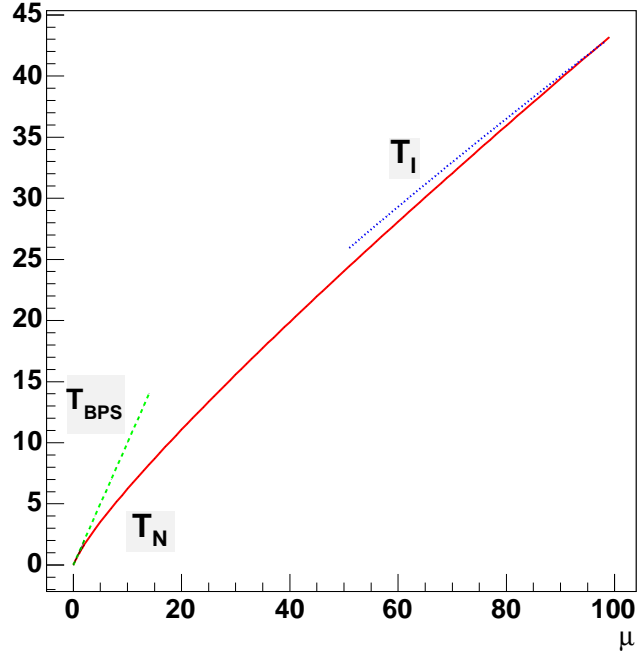,width=100mm,angle=0}}
\caption{Numeric solution for the string tension
$T_N=\frac{g^2}{2\pi m^2}\,T$
as function of the parameter $\mu={m_3\over m}$.
We present the result of numeric computation (the main solid line)
for the tension of $Z_2$ string together with its limiting cases.
The dashdotted line near the origin is the
(normalized and dimensionless) tension of BPS string \rf{bpsten},
(see fig.~\ref{fi:tension23} for more details)
while the dashed
line in the upper right corner
corresponds (again normalized and dimensionless)
to the extreme type I string \rf{lm3ten},
(see again details at fig.~\ref{fi:tension23}).}
\label{fi:tension1}
\end{figure}


\begin{figure}[tp]
\centerline{\epsfig{file=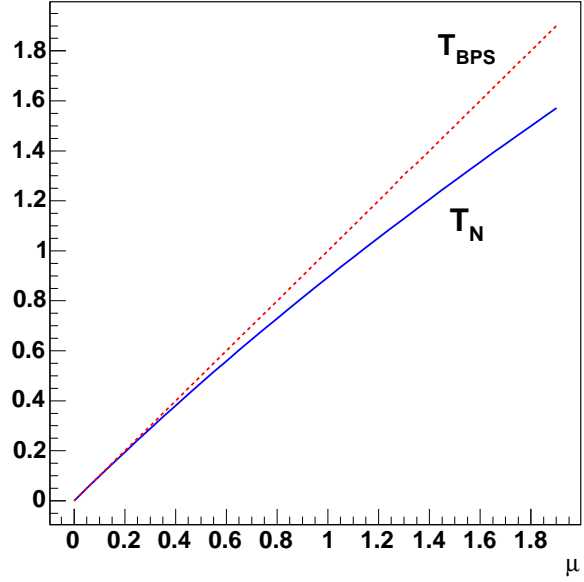,width=90mm,angle=0}}
\centerline{\epsfig{file=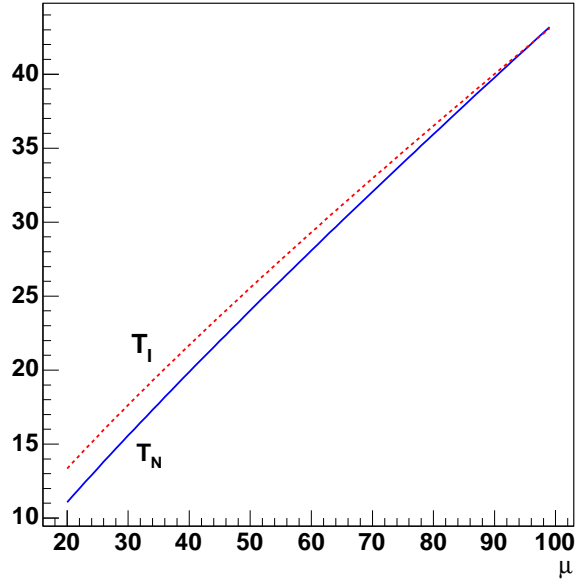,width=90mm,angle=0}}
\caption{
More details for the tensions at small and large values of $\mu$.
The solid line $T_N$ at both pictures
corresponds to the numeric result for the tension of
$Z_2$ string, dashdoted line $T_{BPS}$
at the top picture is the tension of BPS string,
while the dashed line $T_I$
at the bottom picture corresponds to extreme type I string.}
\label{fi:tension23}
\end{figure}

The dependence of the
string tension $T$ (\ref{ten}) on $m_3$ is shown at
fig.~\ref{fi:tension1}, where we plotted dimensionless quantity
$\frac{g^2}{2\pi m^2}\,T$ versus the ratio $\mu = {m_3\over m}$.
The numeric result "interpolates" between the tension of the BPS
string \rf{bpsten} near the origin and the tension (\ref{lm3ten})
of the extreme type I string in the large $m_3$ limit.  Indeed,
we see from fig.~\ref{fi:tension1} (and more details at
fig.~\ref{fi:tension23}) that at small $\mu$ (small $m_3$) the
tension of $Z_2$ string goes very close to the linear dependence
of the BPS string tension of \ntwo QED (\ref{qed}), truncated
according to eq. (\ref{atoxi}). In particular, even at $\mu=1$
the difference of tensions of these two strings is about ten
percent.  Note also that the tension of $Z_2$ string is always
smaller then the BPS bound, in accordance with the expected
\cite{VY} type I superconductivity in \ntwo QED (\ref{qed}).
Moreover, at large $\mu$ or $m_3\gg m$ the tension of the $Z_2$
string in our numeric solution becomes indeed very close to the
tension of extreme ANO string of type I, given by the logarithmic
formula (\ref{lm3ten}); one can see this in the region of large
$\mu ={m_3\over m}$ at fig.~\ref{fi:tension1} and in more details
at fig.~\ref{fi:tension23}.

\section{Non-Abelian strings
\label{ss:nastrings}}
\setcounter{equation}{0}

When $m_3$ approaches $m$ the theory acquires additional symmetry.
In this case the scalar VEV's take the form (\ref{svac}), which respect
the global combined color-flavor symmetry (\ref{c+f}).
On the other hand, the string solution (\ref{sstr}) itself is not invariant
under this symmetry. This means that applying rotation (\ref{c+f})
we generate the whole class of string solutions with the same tension.
In other words the symmetry  (\ref{c+f})  generates orientational
zero modes of the string. Namely, embedding the Abelian
$Z_2$-string (\ref{sstr})
into the non-Abelian theory via (\ref{phi}) and
applying the combined color-flavor rotation, one gets
$$
\Phi_A^a =O\left(
\begin{array}{ccc}
  \frac{g}{\sqrt{2}}\phi & 0 & 0 \\
 0 & \frac{g}{\sqrt{2}}\phi & 0 \\
  0 & 0 & a_0
\end{array}\right)O^{-1} = \frac{g}{\sqrt{2}}\phi  \delta^a_A
+n_a n_A \left(a_0-\frac{g}{\sqrt{2}}\phi\right),
$$
\be
\label{nastr}
A^a_i = n_a\;\frac{\varepsilon_{ij}x_j}{r^2}\, f(r),\ \ \ \ i,j=1,2,
\ee
where we introduced the the unit orientation vector $n_a$ as
\be
n_a=O^a_b\delta^{b3}=O^a_3.
\label{n}
\ee
The solution (\ref{nastr}) interpolates between Abelian
$Z_2$ string
(\ref{sstr}) for which ${\vec n}=(0,0,1)$ and anti-string with
${\vec n}=(0,0,-1)$.

We see that the flux of the string is determined now by an arbitrary vector
$n_a$ in the color space. This makes our string really non-Abelian. As soon
as the symmetry $O(3)_{C+F}$ is not broken in the vacuum, the rotations in
(\ref{nastr}) do not correspond to any gauge generators "eaten" by the Higgs
mechanism, these rotations indeed correspond to the physical orientational
zero modes of the non-Abelian string. To verify this, one can
explicitly construct the {\em gauge invariant} $n_a$-dependent
operators.

As an example, consider the
``non-Abelian" field strength (to be denoted by bold letters)
\cite{SYsu3wall,SYmon},
\be
({\bf{F}}_3^{*})_{AB} =\frac{2}{m^2}\, \varepsilon^{abc}
\bar{\Phi}_A^a F_3^{*b}\Phi^c_B
\,,
\label{gidefi}
\ee
where the subscript 3 marks the $z$-axis, of direction of the string,
and $F_{3}^{*a}=\half\varepsilon_{ij}F^a_{ij}$;
it is by definition gauge invariant.
Eq.~(\ref{nastr}) implies that on string solution
\be
({\bf{F}}_3^{*})_{AB} =\varepsilon_{ABc}
 n_c\, \frac{g^2\phi^2}{m^2}\,
\frac1r \, \frac{df}{dr}\,.
\label{ginvF}
\ee
>From this formula we
readily infer the physical meaning of the orientational moduli $\vec n$:
the direction of the flux of
the {\em color}-magnetic field (defined in the
gauge-invariant way, see (\ref{gidefi})) in the flux tube is
determined by $n_a$.  For Abelian  strings
(\ref{sstr})  only the component with $A,B=1,2$ of
the color-magnetic flux does not vanish.

Our analysis in this section is almost parallel to constructions
in ref.~\cite{ABEKY} where non-Abelian strings were found
in \ntwo SQCD with the gauge group $SU(2)\times U(1)$. However, the
essential difference is
that strings of ref. \cite{ABEKY} topologically exist due to
winding around the $U(1)$ factor of the gauge group, more
precisely they are associated with  $\pi_1(SU(2)\times
U(1)/Z_2)=Z$. Therefore, one can say that these strings are
Abelian in the sense of their Abelian topological origin. Still,
they are non-Abelian in the sense of the presence of
orientational zero modes associated with the rotation of the
string flux inside the gauge group.

In contrast, our strings here are associated with $\pi_1(SO(3))=Z_2$.
Therefore they are non-Abelian both in the sense of their non-Abelian
winding and in the sense of the presence of orientational zero modes.

\section{Effective  world sheet theory
\label{ss:ws}}
\setcounter{equation}{0}

In this section we consider the low energy effective theory on the
string world sheet associated with "slow" dynamics of orientational zero modes.
We will find that the
orientational zero modes are effectively
described by two-dimensional $O(3)$
sigma model. However, unlike the BPS case of
refs.~\cite{HT1,ABEKY,SYmon,HT2}, the sigma model
appearing here is {\em non}-supersymmetric, and this leads to different
physical properties of our four-dimensional theory.

\subsection{ Kinetic term}
\label{kineticterm}

Assume  that the orientational collective coordinates $n_a$
are slow varying functions of the string world-sheet coordinates
$x_k$, $k=0,3$. Then, moduli $n_a$ become the fields on the world sheet
of a two-dimensional sigma model. Since
the vector $n_a$ parametrizes the string zero modes,
there is no potential term in this sigma model. We begin with
the kinetic term (cf. with \cite{ABEKY,SYmon}).

To obtain the kinetic term let us substitute the solution \rf{nastr},
depending upon the moduli $n_a$,
into the action (\ref{su2}) assuming  that
the fields acquire a dependence on the "slow" coordinates $(t,z)$ via
$n_a(t,z)$.
Technically it is convenient to work with the solution
(\ref{nastr}) in the singular gauge, and proceeding to this gauge
we immediately find that we {\em should} modify the solution.

Indeed, the solution (\ref{nastr})
was obtained as an $O(3)_{C+F}$ rotation of the Abelian
string (\ref{sstr}).
Now we make this rotation locally (i.e. depending on $(t,z)$),
therefore, the $(t,z)$ components of gauge potential $(A_0,A_3)$ do not
necessarily vanish, and they
should be added to  the ansatz. This situation is quite familiar
(see e.g.~\cite{ABCMW,SYmeta,ABEKY}), since one routinely encounters it in the
soliton studies.

To begin with, let us rewrite the $O(3)_{C+F}$ rotation in $SU(2)$-terms,
using the representation
\be
O^{ab}=\frac12 \Tr\left(\tau^a U\tau^b U^{-1}\right).
\label{su2rep}
\ee
It is easy to check that vector $n_a$ satisfies the relation
\be
n_a\tau^a=U\tau^3 U^{-1}.
\label{nU}
\ee
Next, we propose an ansatz for the $k=0,3$  components
of the gauge potential
\be
A_k^a\tau^a
=-2i\,  \left( \ptl_k U\right) \,U^{-1}\,\rho (r)\, , \qquad k=0,
3\,,
\label{An}
\ee
where a new profile function $\rho (r)$ (depending only upon the
transverse radius $r=\sqrt{x_1^2+x_2^2}$) is
introduced. In terms of the moduli fields $n_a$ one can write
\be
A_k^a = -
\varepsilon^{abc}\, n_b\, \partial_k n_c \,\rho(r)\,.
\label{Anad}
\ee
using a particular choice for the matrix $U\in SU(2)$
(satisfying $\Tr \left(\tau^3 U^{-1}\d_k U\right)=0$), which
is not uniquely defined by a vector ${\vec n}\in SU(2)/U(1)$
\cite{SYmon}.

The profile function $\rho (r)$ in (\ref{Anad}) is
determined  through a minimization procedure (cf.
\cite{ABCMW,SYmeta}) which generates a separate equation of
motion for $\rho$.
The kinetic term for $n_a$ comes from the gauge and matter kinetic terms
in (\ref{su2}). Using (\ref{nastr}) and (\ref{Anad}), we find
for the $k=0,3$, $i=1,2$ components of the $SU(2)$ gauge field strength
\be
\label{Fni}
F_{ki}=\frac{\tau^a}{2}\left(\ptl_k n_a
\,\frac{\varepsilon_{ij}x_j}{r^2}\,f
\left(1-\rho \right)+
\varepsilon^{abc}n_b\ptl_k n_c
\,\frac{x_i}{r}\,\, \frac{d\rho}{dr}\right).
\ee
We see that in order to get a finite
contribution from $\Tr\ F_{ki}^2$ in the action one has to impose
\be \rho (0)=1
\label{bcfzero}
\ee
and also $\rho (r)$ should vanish at infinity
\be
\rho (\infty)=0.
\label{bcfinfty}
\ee
Substituting (\ref{Fni}) into the action
(\ref{su2}) and including, in addition, the kinetic term for adjoint
matter, we arrive at
\be
S^{(1+1)}= \frac{ \beta}{2}\,   \int d t\, dz
\, \left(\ptl_k\,  n_a\right)^2\,,
\label{o3}
\ee
where the sigma-model coupling
constant $\beta$ is given by the normalization integral
\be
\beta =
\frac{2\pi}{g^2}\,  \int_0^{\infty}
rdr\left[\left(\frac{d\rho}{dr}\right)^2
+\frac{1}{r^2}\, f^2\,\left(1-\rho \right)^2 +
\right. \\
+ \left.  2\left(a_0(1-\rho)-\frac{g}{\sqrt{2}}\phi\right)^2
+2\left(a_0-\frac{g}{\sqrt{2}}\phi(1-\rho)\right)^2
\right].
\label{beta}
\ee

The functional (\ref{beta}) must be minimized with respect to
$\rho$ with the boundary conditions given by
(\ref{bcfinfty}), (\ref{bcfzero}). Varying (\ref{beta}) with respect to
$\rho$
one readily obtains the second-order equation which
the function $\rho$ must satisfy,
\be
-\frac{d^2\rho}{dr^2} -\frac1r\, \frac{d\rho}{dr}
-\frac{1}{r^2}\, f^2 \left(1-\rho\right)
+
2\rho\left(a_0^2+\frac{g^2}{2}\phi^2\right)-
2\left(a_0-\frac{g}{\sqrt{2}}\phi\right)^2=0 .
\label{rhoeq}
\ee
Taking into account this equation we rewrite the sigma model
coupling (\ref{beta}) as
\be
\beta= \frac{2\pi}{g^2}\,I,
\label{betaI}
\ee
where normalization integral $I$ is given by
\be
I=2\int^{\infty}_{0} r d r\;\left[\left(a_0- \frac{g}{\sqrt{2}}\phi\right)^2
+\sqrt{2}g\rho\phi a_0\right].
\label{I}
\ee
We see that
the two-dimensional coupling constant is determined by the
four-dimensional non-Abelian coupling.  This is quite similar
to what was observed for BPS non-Abelian string in \cite{SYmon}.

\begin{figure}[tb]
\epsfysize=9cm
\centerline{\epsfbox{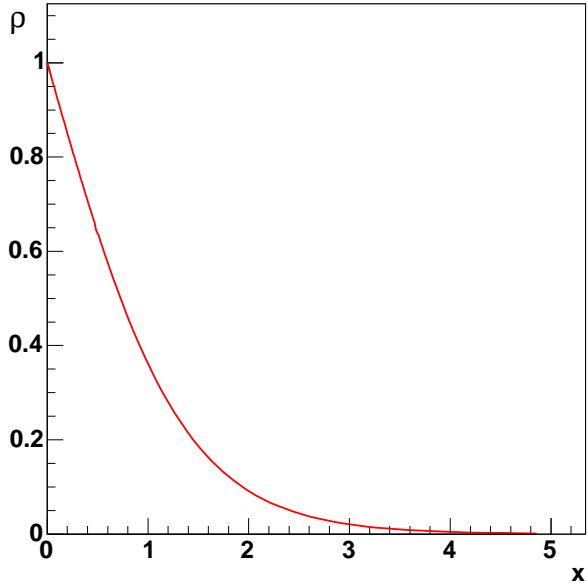}}
\caption{Numeric solution for the profile function $\rho$.}
\label{fi:rho}
\end{figure}

We solve eq. (\ref{rhoeq}) numerically using the numeric solutions
for the string profile functions, see Sect.~\ref{ss:numeric}. The result
for $\rho$ is shown at fig.~\ref{fi:rho}. Note that the first
derivative of $\rho$ does not vanish at $r=0$ similar to the BPS
case, see \cite{ABEKY}. Substituting the solution for $\rho$
in (\ref{I}) we find
\be
I\simeq 0.78 
\label{Ivalue}
\ee
Note that in the BPS case the similar normalization integral
is equal to one \cite{SYmon}. The explanation of this
is related to equality of the monopole mass on the Coulomb branch
to the one
 in the confinement phase present for the case of BPS strings
\cite{SYmon,HT2}. In the theory at hand we don't have such an
equality  as we will explain in Sect. 6.


In summary,  the effective world-sheet theory describing dynamics
of  the string orientational zero modes is
$O(3)$ sigma model.  The symmetry
of this model reflects the presence of the  global $O(3)_{C+F}$
symmetry in the microscopic theory.
The coupling constant of this sigma model is
determined by  minimization of the action  (\ref{beta}) for the
profile function $\rho$. The minimal value of $I$ is given by
(\ref{Ivalue}).

Clearly, the action (\ref{o3}) describes the low-energy limit.
In principle, the zero-mode interaction has higher derivative
corrections which run in powers of
\be
\frac{\ptl}{m} \,,
\label{hd}
\ee
where $m$ gives the order of magnitude of
masses of the
gauge/matter multiplets in our microscopic $O(3)$  theory.
The sigma model (\ref{o3}) is adequate at scales
below $m$ where higher-derivative corrections are negligibly small.

The  same scale $m^{-1}$
determines   the   thickness of the strings we deal with.
In other words,
the effective sigma model action (\ref{o3}) is applicable
at the length scales above $m^{-1}$,
which, thus,  plays the role of an  ultraviolet cutoff for the
model (\ref{o3}).

\subsection{$m_3\neq m$
\label{ss:m3nm}}

To facilitate contact between the microscopic and macroscopic theories,
it is instructive to start from a deformed  microscopic theory
so that the string orientational
moduli are lifted already at the classical level.
Thus, let us drop the assumption $m_3=m$ and introduce a
small mass difference. We will  assume that
\be
\Delta m \equiv \left| m_3-m\right| \ll m\, .
\ee
At $m_3\neq m$ the flavor (global) $O(3)$ symmetry
of the microscopic theory is explicitly broken
down to $O(2) \cong U(1)$ (corresponding to rotations around the third
axis in the $SU(2)/U(1)$ coset space).  Correspondingly,
the moduli of the non-Abelian string are lifted, i.e. the
sigma-model acquires some mass terms.

Let us derive the effective two-dimensional world sheet theory
for the case of $m_3\neq m$. As already discussed in Sect.~\ref{ss:astringsz2},
the solutions of the equations of motion (\ref{streq})
with the minimal windings $n=\pm 1$ directly correspond to the $Z_2$
string and anti-string. Note that both equations of
motion (\ref{streq}) and boundary conditions (\ref{bc})
are written for arbitrary $m$ and $m_3$;
the solution for the  $Z_2$ string is written in terms of the profile
functions in (\ref{sstr}),
while the solution for the $Z_2$ anti-string can be obtained
from (\ref{sstr}) by $ f\to -f $.

At  small $\Delta m $ one can still introduce the orientational
quasi-moduli $n_a$.  In terms of the effective two-dimensional
world sheet theory $\Delta m\neq 0$ leads to a shallow potential
for the quasi-moduli  $n_a$. The two minima of the
potential  at $n=\{0,0,\pm 1\}$ would correspond directly
to the $Z_2$ string and anti-string.

Let us derive this potential in the leading approximation in $\Delta
m$, this can be done just by substituting
the string solution (\ref{nastr}) into
the potential in the action (\ref{su2}). Keeping the linear
in $\Delta m$ terms, one gets
\be
S^{(1+1)}= \frac{ \beta}{2}\,
\int d t\, dz \,\left( \left(\ptl_k\,  n_a\right)^2\, +\gamma m_0
\Delta m(1-n^2_3) + {\rm const}\right),
\label{mo3}
\ee
where
$m_0=\half (m+m_3)$ while the constant $\gamma$ is given by the
ratio
\be
\gamma=\frac{I_1}{I}\simeq 0.8,
\label{gamma}
\ee
where $I$ is the normalization integral (\ref{I}), with the numeric
value (\ref{Ivalue}), while
\be
I_1=4\int_0^{\infty} rdr \left|a_0^2-\frac{g^2}{2}\phi^2\right|
\label{I1}
\ee
and after substituting the numeric results for the profile functions
(see Sect.~\ref{ss:numeric}) into (\ref{I1}) one finds
\be
I_1\simeq 0.63 
\label{I1cnum}
\ee
Note, that in linear in $\Delta m$ approximation we do not
need to modify the string solution (\ref{nastr}), since, as usually,
corrections to the string solution contribute only in the
quadratic in  $\Delta m$ order, due to equations of motion,
and in this approximation,
the integral \rf{I1} was calculated, putting $m_3=m$.

To extract the $(\Delta m)^2$ corrections to the potential in
(\ref{mo3}) one has to modify the string solution introducing
a new profile function like it is done in \cite{SYmon} for the
case of BPS string. We have checked that already in $(\Delta m)^2$ order
the potential has a complicated $n_3$-dependence:
generally it contains an infinite
series in powers of $(1-n^2_3)$.

The theory in (\ref{mo3}) is massive deformation of the $O(3)$ sigma model.
The potential in (\ref{mo3}) is similar to the one
arising in the case of non-Abelian BPS strings, see \cite{T,SYmon,HT2};
however, the difference is that for the BPS case the potential appears
only at the quadratic order in $\Delta m$ while in \rf{mo3}
it arises already in linear order. The reason is
that in the BPS case the world sheet theory has \ntwo supersymmetry
and the (twisted) superpotential is proportional to $\Delta m$,
that ensures the bosonic potential  proportional to  $(\Delta m)^2$.
Note, that the non-analytic dependence of the
potential in (\ref{mo3}) on the mass difference $\Delta m= \left|
m_3-m\right|$ cannot appear in supersymmetric sigma model.

\subsection{Fermionic zero modes}

As we already explained in Sect.~\ref{ss:model} in the \ntwo limit of small
$m_3$
our string solution is 1/2-BPS saturated. This means that,
four charges (out of the eight supercharges of the four-dimensional \ntwo
algebra) act on the string solution
(\ref{bpsstr}) trivially, while the remaining four charges generate the
four fermionic
zero modes (called supertranslational, since they are
superpartners to the two translational bosonic zero
modes). The corresponding four fermionic moduli
are superpartners to the coordinates $(x^{(0)}_1,x^{(0)}_2)$
of the "string center", and the four SUSY charges, vanishing on the BPS
string solution, turn (after we consider moduli as slow functions
of the world sheet co-ordinates $(z,t)$) into the four generators of the
two-dimensional
\N2 supersymmetry algebra; the  supertranslational
fermionic zero modes for the $U(1)$ ANO string in
were studied in \cite{VY}.

As we increase $m_3$, the \ntwo supersymmetry breaks down to
\none, and the string solution is no longer BPS-saturated. Still the number of
the fermionic zero modes does not jump: all four generators of the \none
SUSY algebra act now non-trivially on the solution and, therefore, produce
the same number of the "supertranslational"
(do not have, in fact, any supersymmetry in the world sheet
theory at $m_3 \neq 0$)
fermionic zero modes \cite{VY}.
These "supertranslational" modes decouple from the internal dynamics
and they are not essential for our purposes below.

Approaching the point $m_3=m$, our theory
acquires an  additional symmetry (\ref{c+f}),
responsible for appearance of orientational zero modes
of the string, described by the
$O(3)$ sigma model (\ref{o3}). As we already mentioned, one should
not expect world sheet supersymmetry at $m_3 \neq 0$, and
nothing special happens at the point $m_3=m$: supersymmetry
 is still absent in the
world sheet theory.
Therefore, we do not expect any "superorientational" fermionic
zero modes to appear at the point $m_3=m$, and we expect that the
internal string dynamics is described by the {\em non-supersymmetric} version
of
the $O(3)$ sigma model (\ref{o3}). This is extremely important for the
physical conclusions of the Sect.~\ref{ss:dynamics} below.

Note, that the properties of the world sheet theory we consider are
essentially different from the case of non-Abelian strings, considered
in \cite{ABEKY,SYmon}. For
that non-Abelian BPS string four supercharges act trivially on the
string solution and turn into the SUSY generators
in the effective world sheet theory. The four
"superorientational" fermionic zero modes were found in \cite{SYmon},
the corresponding four fermionic moduli are
superpartners of bosonic variables  $n_a$ in the effective
world sheet \ntwo $O(3)$ sigma model. It is also useful to notice, that
the world-sheet supersymmetry imposes serious restrictions to the possible
form of the potential in the massive deformation of the sigma-model, while
our conclusions of Sect.~\ref{ss:m3nm} rather point out that the
potential  in (\ref{mo3}) ( as well as generic form of
corrections arising beyond the linear
approximation in $\Delta m$) is inconsistent with the world-sheet
supersymmetry.

\section{World sheet dynamics
\label{ss:dynamics}}
\setcounter{equation}{0}

In the previous section we show that the internal world sheet
dynamics of our non-Abelian string is described in terms of the
(massive deformation of) $O(3)$
sigma model. Although this theory is asymptotically free and
runs into strong coupling regime its physics is well understood.
In this section we review some known results about $O(3)$
sigma model and interpret them in terms of strings in
four-dimensional \1N$^{*}$ theory.

\subsection{Quasiclassical limit}

 Let us start with the case of massive  $O(3)$ sigma model
at $m_3\neq m$.
The sigma model (\ref{mo3})
is asymptotically free \cite{Po3}; at large distances (low
energies) it gets into the strong coupling regime.  The  running
coupling constant  as a function of
the energy scale $E$ at one-loop is given by


\begin{figure}[tp]
\centerline{\epsfig{file=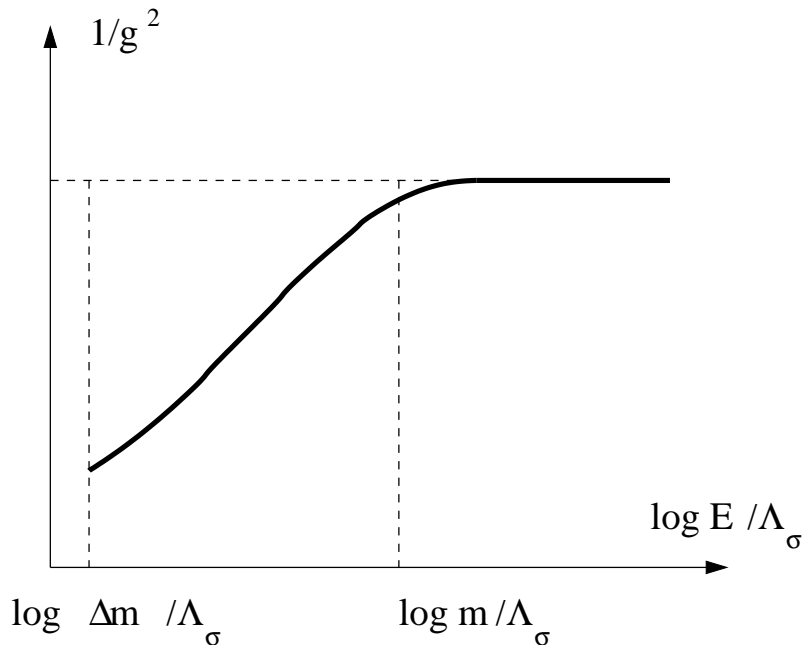,width=105mm,angle=0}}
\caption{Beta-function of four-dimensional (microscopic) and two-dimensional
(macroscopic) theory. At energies $E \gg m$ the coupling does not run due to
conformal invariance of \4N theory and equals to some bare value $g_0$.
At the scales below $m$ the running of coupling
is determined by two-dimensional world-sheet theory and goes towards the
strong-coupling regime.}
\label{fi:beta}
\end{figure}

\be
4\pi \beta = 4\ln {\left(\frac{E}{\Lambda_{\sigma}}\right)}
+\cdots,
\label{sigmacoup}
\ee
where $\Lambda_{\sigma}$ is the dynamical scale of the sigma
model. As was mentioned previously,
the ultraviolet cut-off of the sigma model at hand
is determined by  $m$. At this UV cut-off scale
Eq.~(\ref{betaI}) holds.
Hence,
\be
\Lambda^4_{\sigma} = m^4 e^{-\frac{8\pi^2}{g^2}\,I} .
\label{lambdasig}
\ee
Note that in the microscopic theory, due to the VEV's of
the squark fields, the coupling constant is frozen at
$m$. There are no logarithms in microscopic theory
neither above this
scale (because of conformal invariance of \nfour theory)
nor below it, however below $m$ the logarithms of the macroscopic theory
take over, see fig.~\ref{fi:beta}.

Consider, first, the  $O(3)$ sigma
model in the quasiclassical regime of large $\Delta m$; to be
more precise take  $\Delta m\gg \Lambda_{\sigma}$
(note that we always assume that $\Delta m\ll m$).
The coupling constant of the $O(3)$ sigma model is frozen
at  $\Delta m$. Therefore in the region of large $\Delta m$,
$\Delta m\gg \Lambda_{\sigma}$  the inverse coupling $\beta$ is
large  and we can study the model in the quasiclassical
approximation.

\begin{figure}[tp]
\centerline{\epsfig{file=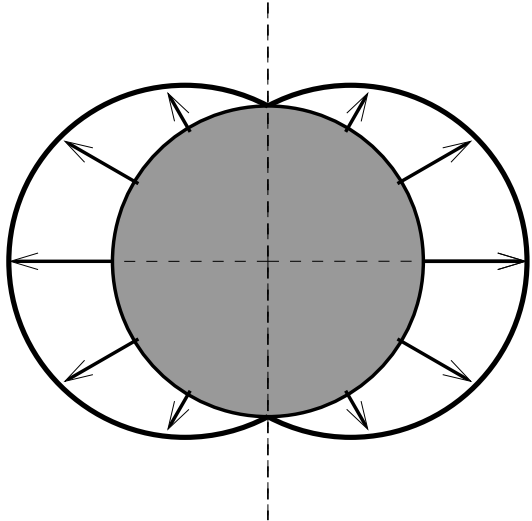,width=70mm,angle=0}}
\caption{The meridian slice of the target space sphere
${\vec n}\in S^2$ of the
$O(3)$ sigma-model.  Arrows present the scalar potential in
(\ref{mo3}), with their lengths being the strength of the
potential. Two vacua of the model are seen
at the north and south poles $n_3 = \pm 1$ of the sphere.}
\label{fi:pot}
\end{figure}

The potential  of the $O(3)$ sigma model  is
derived in Sect.~\ref{ss:m3nm}  and the
result is schematically presented at
fig.~\ref{fi:pot}. This potential is very similar
to the potential in the case of BPS non-Abelian string, and the
analysis below will be rather close to those presented in
\cite{T,SYmon,HT2}.


The potential has two vacua at the north ($n_3= 1$)
and south ($n_3=- 1$) poles. These vacua correspond to Abelian
$Z_2$-string and anti-string, see (\ref{str}).
There is a sigma model kink (domain wall in the world sheet theory),
interpolating between
two vacua of massive $O(3)$ sigma model. This kink is interpreted
as a monopole which provides a junction of the $Z_2$-string and
anti-string. Remember that monopole has magnetic charge one
while $Z_2$-string and anti-string have charges $\pm 1/2$;
therefore the string cannot end on a monopole (and this is a
reason for the stability of $Z_2$-strings), and a
monopole can appear only as a junction of $Z_2$-string and anti-string.
Note also, that in the Higgs vacuum monopoles do not exist as a
free states. They are in the confinement phase, attached to
ANO strings carrying their magnetic fluxes.

The similar identification of a two-dimensional sigma model
kink as confined four-dimensio-\-nal monopole was presented in
refs.~\cite{T,SYmon,HT2} for the case of BPS monopoles,
and a lot of arguments was presented in favor of this interpretation.
In particular, in \cite{SYmon} the first order equations for the
1/4-BPS string junction were explicitly solved and the solution was shown
to be determined in terms of the kink solution of
massive $O(3)$ sigma model. Moreover, the masses of four-dimensional
monopole and two-dimensional kink were shown to coincide for these
solutions
\footnote{Recently a general solution for 1/4-BPS junctions of
semilocal strings was obtained in \cite{INOS}.}.

Here we do not have explicit analytic solutions, but we will use this
identification to calculate the mass of monopole in the confinement phase.
First, rewriting the action (\ref{mo3}) in the holomorphic
representation upon stereographic projection, one gets
\be
S^{(1+1)}= 2\beta\,   \int d t\, dz
\,\frac{\left|\ptl_k\,  w\right|^2\,
+\gamma m_0 \Delta m|w|^2}{(1+|w|^2)^2},
\label{mo3w}
\ee
where $w$ is a complex field. Two classical vacua are now located
at $w=0$ (north pole) and $w=\infty$ (south pole).
Note that the model (\ref{mo3w}) has $U(1)$ symmetry
$w\to e^{i\theta} w$, since the potential of massive deformation
of the sigma model (\ref{mo3}), as we already mentioned,
does not break $O(3)_{C+F}$ completely, leaving
$U(1)$-rotations do not acting on $n_3$.

The kink solution, interpolating between two
classical vacua is now easy to find. It is
a static solution depending only on the coordinate $z$. In
one-dimensional case the equations of motion can be always integrated
by energy conservation law, and for
the kink in (\ref{mo3w}) we get the first-order equation
\be
\ptl_z w=-\sqrt{\gamma m_0\Delta m} \,w
\label{kinkeq}
\ee
with the solution, given by
\be
w (z) = \exp\left(-\sqrt{\gamma m_0 \Delta m}\, (z-z_0) -i\theta\right).
\label{kink}
\ee
Here $z_0$ is the center of kink, while $\theta$ is an arbitrary
phase; in fact, these two parameters enter only in the
combination $\sqrt{\gamma m_0 \Delta m}z_0 -i\theta$, so that
the notion of the kink center gets complexified.

Now, substituting the kink solution into the action
(\ref{mo3w}), one gets the mass of the kink
\be
M_{\rm kink}=2\beta \sqrt{\gamma m_0 \Delta m}.
\label{kinkmass}
\ee
As we already explained, the kink mass (\ref{kinkmass}) should
be identified with the mass of the confined monopole providing
a junction of $Z_2$-string and anti-string.
Expressing
the kink mass in terms of parameters of four-dimensional
theory, using (\ref{betaI}) and (\ref{gamma}), one gets
the  mass of confined monopole
\be
M_{\rm m}^{\rm conf}
=\frac{4\pi}{g^2}
\sqrt{m\Delta m\,I_1 I}\left(1+O\left(\frac{\Delta m}{m}\right)\right).
\label{monmass}
\ee
with the numeric coefficient $\sqrt{I_1I} \simeq 0.7$ (see \rf{Ivalue}
and \rf{I1cnum}).
On the Coulomb branch at $m_3=0$, the monopole mass
is given by the Seiberg-Witten formula, which in the weak coupling
regime reads
\be
M_{\rm m}^{\rm Coulomb}
=\frac{4\pi}{g^2}\,m.
\label{monmassc}
\ee
Comparing two results  (\ref{monmass})  and (\ref{monmassc})
for the monopole masses, one finds that when moving into the Higgs phase
by increasing $m_3$, the monopole confines and its mass decreases:
given at small $\Delta m$ by (\ref{monmass}).
Classically, in the non-Abelian point $m_3=m$ it vanishes,
while the monopole size becomes infinite; however, in the
next section we find, that this does not happen in quantum
theory, due to the (two-dimensional) non-perturbative instanton
effects on the string world sheet.

To conclude this section, let us discuss the
physical meaning of extra modulus $\theta$ of the kink solution \rf{kink}.
There is a continuous family of solitons, interpolating between the north and
south poles of the target space sphere, see fig.~\ref{fi:pot}.
The soliton trajectory can follow any meridian,
due to unbroken $U(1)$ symmetry. After quantization,
the phase $\theta$ leads to the whole tower of
"dyonic" states \cite{Dorey,Losev} with the non-zero
charge with respect to unbroken $U(1)$ symmetry.

\subsection{Quantum limit}

In this section we consider the quantum theory at $\Delta m\to0$.
As we already discussed, the non-Abelian string in \1N$^{*}$ theory
is a {\em non} BPS string, and therefore its world sheet dynamics
is described in terms of the {\em non} supersymmetric $O(3)$
sigma model (\ref{o3}).
Now, let us review the known results on the $O(3)$ sigma model
and interpret them in terms of strings and monopoles in our
original four dimensional theory.

The exact spectrum of the
non-supersymmetric $O(3)$ sigma  model was found long ago in
\cite{ZZ}. Moreover, it was shown in ref.~\cite{W1/N}
that the main features of the exact solution can be
qualitatively understood in terms of the $1/N$ expansion of $CP^{N-1}$
models, if one formally puts $N=2$ \footnote{The same approach
works for the \ntwo supersymmetric $O(3)$ sigma model
\cite{A-GF}, in the supersymmetric case the exact solution
\cite{ShW} (and even more recent results based on mirror
symmetry \cite{HoVa}) can be also qualitatively understood in terms
of the $1/N$ expansion \cite{W1/N}.}.
It turns out that the dynamics of the
$O(3)$ sigma models for supersymmetric and non-supersymmetric
cases are rather
different. Below we remind these differences, which lead to
quite different properties of the non-Abelian non-BPS
strings from the BPS strings studied in \cite{SYmon}.

In the $CP^1$ formulation the action of the $O(3)$ sigma model (\ref{o3})
reads
\be
S_{O(3)}=\frac{\beta}{2} \int dt dz |{\cal D}_k Z_l|^2,
\label{cp1}
\ee
where $Z_i$ is the complex doublet of $SU(2)$, constrained by
$\sum_l|Z_l|^2=1$,
while the covariant derivative
${\cal D}_k=\ptl_k -i{\cal A}_k$ contains a
two dimensional non-dynamical $U(1)$ gauge field ${\cal A}_k$.
This is a particular case of the
$CP^{N-1}$ model, where in the action
(\ref{cp1}) $Z_l$ should be treated as $N$-component
complex field in the fundamental representation of $SU(N)$.
The $1/N$-expansion yields two main results \cite{W1/N}: the
field $Z_i$ acquires a mass of order
of $\Lambda_{\sigma}$, and the $U(1)$ gauge field ${\cal A}_k$
acquires at one loop level the standard kinetic term.

Classically the vector $n_a$ can be directed anywhere on the
sphere $S^2$, so one naively expects the spontaneous breaking of the
$SU(2)$ symmetry
and the massless Goldstone modes. However,
the first of the observations above shows, that the spontaneous
$SU(2)$ symmetry breaking  does not occur in quantum theory and
there are no massless particles. The second result implies
confinement for $Z_l$ fields \cite{W1/N}. The reason is
that the Coulomb potential between two charges in two dimensions
is linear in the distance between these charges
\be
V_{\rm Coulomb}^{\rm 2d}=\Lambda^2_{\sigma}|z_1-z_2|,
\label{2Dconf}
\ee
where $z_1$ and $z_2$ are positions of the charges.
In fact, the exact spectrum of $O(3)$ sigma model \cite{ZZ}
contains one triplet, which is interpreted as
$\bar{Z}Z$-meson in the adjoint representation of global $SU(2)$.

Before interpreting these results in terms of strings and
monopoles in four dimensional theory, let us remind the similar
results for the
\ntwo supersymmetric $O(3)$ sigma model \cite{W1/N}.
The presence of the world-sheet fermions
drastically changers the above outlined
picture. The fermions generate a mass of the order of  $\Lambda_{\sigma}$
for the $U(1)$ gauge field via the anomalous one loop diagram.
Thus, the linear potential (\ref{2Dconf}) is screened and there is
no confinement, in accordance with the exact spectrum of
\ntwo $O(3)$ sigma model, which contains doublets of $SU(2)$,
interpreted as massive $Z_l$ particles.

In fact, in the \ntwo $O(3)$ sigma model there are two vacua,
and this is easy to understand. One can start with massive
deformation of the sigma model in quasiclassical regime, like we did
in previous section (cf. also with \cite{Dorey,SYmon}).
Then two vacua are just the north and south poles at fig.~\ref{fi:pot},
corresponding, as we know,
to two elementary  Abelian strings
with minimal windings in four dimensions.
When we reduce the mass parameter $\Delta m$ the number of vacua
in supersymmetric version of the model does not change due to the Witten
index \cite{Windex}. Thus, in massless case we still have
two quantum vacua in the supersymmetric sigma model, to be
interpreted as two
elementary non-Abelian strings \cite{SYmon}. The order parameter
to distinguish between these two vacua is the bifermionic condensate
generated by instantons \cite{NSVZ}.

As soon as we have two vacua at $\Delta m=0$
we still have a kink (in fact two
kinks \cite{HoVa}) interpolating between them. This kink is
interpreted as a quantum version of the non-Abelian monopole,
providing 1/4-BPS string junction between two elementary
strings. Classically this monopole would have infinite size
and zero mass, but this does not  happen in quantum theory.
There is no massless states in $O(3)$ sigma model, therefore
the kink/monopole has finite size (of the order of $\Lambda_{\sigma}^{-1}$)
and non-vanishing mass (of the order of $\Lambda_{\sigma}$),
given by the anomalous term in \ntwo superalgebra~\cite{SYmon}.
The kink
of \ntwo $O(3)$ sigma model is a doublet of global $SU(2)$ and
can be interpreted as a $Z_l$-particle in the $CP^1$ formulation
(\ref{cp1}) \cite{W1/N}, see also \cite{HoVa}. Thus, we conclude
that confined non-Abelian BPS monopoles are associated with the
$Z_l$ fields in the two-dimensional world sheet $CP^1$ sigma
model.

Now let us return to our case of the non BPS string,
described by non supersymmetric $O(3)$ sigma model.  First,
in the $O(3)$ sigma model there
is the only vacuum state, and therefore
the only non-Abelian string with minimal tension in four dimensional
theory. Second, as soon as there are no different vacuum states
in the world-sheet theory, there is no room for the kinks
interpolating between them. Hence, there is {\em no}
non-Abelian monopoles in the limit $\Delta m=0$; where do they
disappear?

First, the absence of two elementary strings is easy to understand.
Due to mixing, two vacua of the $O(3)$ sigma model present
at large $\Delta m$ are split, and in the limit $\Delta m=0$
we end up
with two energy levels with the energy gap of the order of
$\Lambda_{\sigma}$\footnote{This splitting does not occur in
SUSY case, since vacuum energies are protected by supersymmetry
from quantum corrections, and the number of vacuum states does
not change in accordance with the Witten index.}.
Thus we do have
the {\em only one} non-Abelian string with minimal tension, the other
tension is larger. Note, that the splitting of the order of
$\sim \Lambda^2_{\sigma}$ is rather small, compare to the
classical tension of each string ($\sim m^2/g^2$).

\begin{figure}[tp]
\centerline{\epsfig{file=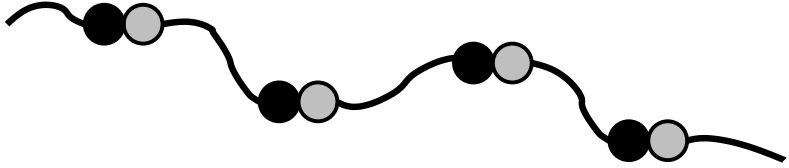,width=105mm,angle=0}}
\caption{Two-dimensional confinement of monopoles (black dots) and
anti-monopoles (grey dots) on the string.}
\label{fi:2dconf}
\end{figure}

Still, one can have junction of the "minimal" string
with the exited string, and this junction should be again interpreted
as a non-Abelian monopole. However, since the exited string carries
more energy, one would expect that a monopole comes together
with an anti-monopole in a meson-like configuration,
see fig.~\ref{fi:2dconf}.
The total energy of such object is determined by length of
the exited string between monopole and anti-monopole
times the tension difference between two lightest strings.
This gives the energy
$ \sim \Lambda^2_{\sigma}|z_1-z_2|$,
where $z_1$ and $z_2$ are positions of monopole and
anti-monopole, or  we immediately
recognize the confining potential \rf{2Dconf}.


As we already
mentioned, the $O(3)$ sigma model kink is interpreted as
a $Z_l$-particle in the $CP^1$ description. In (non supersymmetric)
$O(3)$ sigma model the $Z_l$-particles are confined. Now we see,
that the above conclusion that monopole and anti-monopole
form a meson-like configuration on the string (see fig.~\ref{fi:2dconf})
is in a complete accordance with known results about the confinement of
$Z_l$-particles/kinks in the $O(3)$ sigma model. 

Since kinks (or the $Z_l$-particles) of the $O(3)$ sigma model 
are in the doublet representation of the
global unbroken $SU(2)_{C+F}$, the kink-anti-kink meson can be either
scalar or triplet of $SU(2)_{C+F}$. Clearly, the scalar is unstable,
because kink and anti-kink with the "opposite" quantum numbers 
can annihilate each other. However,
they cannot annihilate each other when forming a triplet state and,
as we already mentioned, the exact spectrum of $O(3)$ model indeed
consists of the stable massive triplets of $SU(2)_{C+F}$ \cite{ZZ}.
In four dimensions these states correspond to the (attached to the string) 
stable monopole-anti-monopole mesons in the triplet representation of the 
unbroken $SU(2)_{C+F}$.

To avoid confusion we would like to stress that monopoles
are in the confinement phase in the Higgs vacuum we consider in
this paper. This four-dimensional confinement implies that
monopoles do not
exist as free states, but they are attached to strings carrying
magnetic flux. We see now, that on top of this four-dimensional
confinement, we have an "extra"
two-dimensional confinement in the $O(3)$ sigma model, which
additionally forces monopoles and anti-monopoles
to form a meson-like configuration
on the string they are attached to (see fig.~\ref{fi:2dconf}).

\section{Conclusions
\label{ss:conclusion}}

In this paper we presented a simplest model for the non-Abelian
strings. We have considered the \4N supersymmetric $SU(2)$ gauge theory,
perturbed by the mass terms for the chiral multiplets, in the Higgs
vacuum. We have shown that at the special value of mass parameters
$m_3=m$ the theory
has an unbroken global $O(3)_{C+F}$ symmetry,
responsible for the appearance of orientational zero modes
of the $Z_2$ strings, associated with the rotation of their color fluxes
inside the $SU(2)$ group. The presence of these zero modes makes
the $Z_2$ strings non-Abelian.

We have worked out the effective theory on the world sheet of the
non-Abelian string. It turn out to be the two-dimensional
(non-supersymmetric) $O(3)$ sigma model.
Then we have translated the results for the $O(3)$ sigma model
to the physics of strings in four dimensions.
Like in \cite{T,SYmon,HT2}, the confined 't Hooft-Polyakov monopole
is identified with the
kink of two-dimensional $O(3)$ sigma model, and this allowed us to
calculate its mass. In
particular, we demonstrated that besides the four dimensional
confinement, which ensures
that monopole is a $Z_2$-string junction, the monopoles are also
confined in the two-dimensional sense.
Namely, monopoles and anti-monopoles form meson-like configuration
on the string they are attached to. The spectrum of the theory contains
these stable monopole-anti-monopole mesons in the triplet
representation of $O(3)_{C+F}$.

Since there are no experimental signs for "Abelization"
in the real world QCD, it is plausible to expect that the
non-Abelian strings, like we discussed here and in \cite{HT1,ABEKY},
can be also responsible for the confinement there.
The orientational modes of the
non-Abelian string appear due to
the presence of unbroken global color-flavor symmetry
($O(3)_{C+F}$ in the model considered above) and
ensure the "de-Abelization" of the string.

All results of this paper concern the Higgs vacuum at
weak coupling, i.e. in the Higgs vacuum the color electric
(adjoint) charges condense, giving rise to the non-Abelian magnetic
strings and confinement of monopoles. However, it is well known
that  \1N$^{*}$ theory has the $SL(2,Z)$ duality group \cite{MO,DW}.
In particular, the S-duality transformation was analyzed in \cite{DW}
at small $m_3$,
and it was shown that it exchanges the Higgs and monopole vacua. Thus,
the confinement of monopoles in the Higgs vacuum is mapped to
the confinement of W-bosons and quarks (if we introduce quarks)
in the monopole vacuum.

Can we extrapolate this picture to large
$m_3$? In particular, can we get the non-Abelian electric
$Z_2$-strings in the monopole vacuum by S-duality transformation
from the non-Abelian magnetic $Z_2$-strings in the Higgs vacuum?
Unfortunately, this is hard to do. The duality transformations at
arbitrary $m_3$ were studied in \cite{Do,DoK}, and it was shown that
BPS-data like chiral condensates and domain wall tensions in the
Higgs and monopole vacua are mapped one into another by S-duality.
However, the strings we found are {\em not} BPS, they become BPS strings
only in the "Abelian" limit of small $m_3$ (and in this limit the
Abelian quark confinement in the monopole vacuum is well understood
\cite{SW1,DS,FG,VY}). Therefore, it hard at the moment
to make any conclusion on
the existence and properties of $Z_2$-strings  in the monopole
vacuum just from S-duality.  Moreover, the S-duality transformation
maps small $g$ to large $1/g$. Thus, say, the Higgs vacuum at
small $g$  is mapped to the monopole vacuum at large bare \nfour
coupling, and this is not what can be analyzed by our methods.
It would be nice to study the
theory, starting from the small bare \nfour coupling (of course
the running coupling at the monopole vacuum is always large).

Let us finally point out, that the $O(3)$ sigma model, which is a world-sheet
theory for the effective string in the perturbed $SU(2)$ \4N theory near the
color-flavor locked vacuum, is similar in certain aspects
to the string "spin" sigma-models, arising in the $SU(2)$ flavor
sector of the
\4N Yang-Mills theory in the context of the AdS/CFT correspondence
\cite{AdS}. At the moment the pictures for the effective and fundamental
string sigma-models are still essentially different: the effective string
sigma-model is not conformal and runs into strong coupling, where quantum
effects change drastically the naive classical picture, while on
the fundamental string sigma-model side only the predictions of
the classical theory can be used in some cases. However, one can
still hope that these two ways, when the world-sheet sigma-models
arise in the supersymmetric Yang-Mills theory, are complementary
realizations of the same phenomenon; the string/gravity
dual of the \1N$^{*}$ theory itself was studied in \cite{PoSt}.

\section*{Acknowledgments}

We are grateful to  A.~Gorsky, M.~Shifman and A.~Vainshtein for very
useful discussions, and to M.~Hindmarsh and M.~Kneipp for valuable
communications. The work of V.~M. was supported in part
by DAAD, Dynasty Foundation and personal grant of the St.~Petersburg Governor.
The work of A.~M. was partially supported by the RFBF grant
No.~02-02-16496, INTAS grant 00-561, the grant for support of scientific schools 1578.2003.2,
and the Russian Science Support Foundation.
The work of A.~Y. was partially supported by the RFBF grant No.~02-02-17115
and by Theoretical Physics Institute at the University of Minnesota.


\end{document}